\documentclass[twocolumn,eqsecnum,showpacs,pra]{revtex4}
\usepackage{amsmath}
\usepackage{amssymb}
\begin{document}
%\draft
\title{ Bures distance as a measure of entanglement for symmetric 
two-mode Gaussian states}
\author{Paulina Marian}  
\author{ Tudor A. Marian}
\affiliation{ Centre for Advanced  Quantum Physics,
University of Bucharest, P.O.Box MG-11,
R-077125 Bucharest-M\u{a}gurele, Romania}
\date{\today}

\begin{abstract}
We evaluate a Gaussian entanglement measure for a symmetric two-mode Gaussian
state of the quantum electromagnetic field in terms of its 
Bures distance to the set of all separable Gaussian states.
  The required minimization procedure was  considerably 
simplified by using
the remarkable properties of the Uhlmann fidelity as well as the standard form II 
of the covariance matrix of a symmetric state. 
Our result for the Gaussian degree of entanglement 
measured by the Bures distance
depends only on the smallest symplectic eigenvalue
 of the covariance matrix of the partially transposed density operator. It is thus
consistent to the exact expression of the entanglement of formation 
for symmetric
two-mode Gaussian states. This non-trivial agreement is specific to the Bures metric.
\end{abstract}

%\twocolumn
\maketitle
\section{Introduction}
In recent years many attempts to quantify the entanglement 
of a Gaussian state have been made due to the  experimental 
interest in using such states in quantum information processing \cite{BL,H}.
Work on  the inseparability properties  of two-mode Gaussian states (TMGSs) 
was also stimulated by the formulation of their separability criterion \cite{Duan,Si1}.  Simon proved that a TMGS is separable if and only if the non-negativity of its density matrix is preserved under 
partial transposition  (PT) \cite{Peres}. Simon has written the PT-criterion in a  Sp $(2,\mathbb{R})  \times$ Sp $(2,\mathbb{R})$ 
invariant form which allows one to easily check whether a two-mode Gaussian state is separable 
or not \cite{PTH}. Denoting by $\tilde{k}_-$ the smallest symplectic eigenvalue 
 of the covariance matrix  (CM) of
the PT-operator $\rho^{PT}$, the separability criterion reads $\tilde{k}_- \geq 1/2$ for a separable TMGS and $\tilde{k}_- < 1/2$ for an entangled one. 

It is interesting to recall that
a computable inseparability measure 
for an arbitrary bipartite state was proposed in Refs. \cite{K,VW}  in terms of {\em negativity} defined as the absolute value of the sum of the negative 
eigenvalues of the PT-density operator. As proved by Vidal and Werner, 
the negativity is an entanglement monotone. For TMGSs the negativity
turned out to be an expression  depending  only on $\tilde{k}_-$ \cite{VW}.

With regard to other accepted measures of entanglement for a TMGS, 
 the only {\em exact} evaluation at present
appears to be  the entanglement of formation 
(EF) for  a symmetric TMGS \cite{G}. The optimal pure-state decomposition required to define the EF  was found in this case to be in terms of Gaussian states. 
 For an arbitrary TMGS, {\em a Gaussian entanglement of formation} 
was further defined 
using its optimal decomposition in pure Gaussian states \cite{W}.
 Following the prescription of Ref.\cite{W}, an evaluation of the Gaussian EF
 for a two-mode squeezed thermal state (STS) was given in Ref.\cite{Jiang}. 
 In the general case an insightful formula for the Gaussian EF was not yet written.   
 One can notice that, for a symmetric TMGS, the amount of entanglement 
is described by monotonous functions 
  (negativity and EF)
  depending on only $\tilde{k}_{-}$.
The situation is different for other special TMGSs. 
A disagreement
 between the Gaussian EF and the negativity 
of the Gaussian states having extremal negativity at fixed global
 and local purities \cite{Aa} was recently noticed in Ref.\cite{AI}.
 
 Following the earlier distance-type proposal for quantifying entanglement due to Vedral and co-workers \cite{Ved},  a class of {\em 
 distance-type Gaussian measures 
of entanglement} could be defined with respect to only the set of separable Gaussian states identified by the separability criterion \cite{Si1}. 
To our knowledge, the first authors who  used and evaluated numerically
 a
 Gaussian measure of entanglement were Scheel and Welsch in Ref.\cite{Scheel}.
In our paper \cite{PTH1} co-authored with  Scutaru, an explicit analytic
Gaussian amount of entanglement was calculated for
 a STS by using the
Bures distance. 
Note that the STSs are important non-symmetric
TMGSs that can be produced experimentally and are used  in the protocols 
for quantum teleportation.
 Interestingly, in the STS-case the Gaussian entanglement measured 
by Bures distance  \cite{PTH1} and
 the Gaussian EF \cite{Jiang}  were found to be in agreement. They 
are monotonous functions of the same parameter \cite{com}  which
 cannot be expressed only in terms of $\tilde{k}_{-}$. 
Therefore, the negativity of a STS is not equivalent to the 
two Gaussian measures of entanglement evaluated at present \cite{PTH1,Jiang}.

Our aim in the present work is to apply the Bures distance
as a measure of entanglement for symmetric two-mode Gaussian states
in the framework of the Gaussian approach and compare 
the result to the exact EF for a symmetric TMGS.
The body of the paper is structured as follows. 
We recall in Sec. II several properties of
 two-mode Gaussian states. 
Here we show that the CM of a symmetric TMGS can be diagonalized by 
the beam-splitter transformation which is orthogonal and symplectic. 
In Sec. III we define a Gaussian amount of entanglement for a symmetric TMGS 
in terms of its 
Bures distance to the set of all separable TMGSs. By using the properties
 of the Uhlmann fidelity between two TMGSs we restrict the reference set of 
all separable TMGSs to its subset of symmetric ones. Application 
of the beam-splitter transformation to both the given inseparable state 
and the set of symmetric separable  TMGSs enables us to evaluate and maximize
in Sec. IV 
just a product of one-mode fidelities. We then give the CM of the 
closest separable state and show that its entries depend on both the 
symplectic eigenvalues of the given CM  and $\tilde{k}_{-}$.
On the contrary, the defined Gaussian amount of entanglement is expressed in 
terms on only $\tilde{k}_{-}$, being thus consistent to the 
exact evaluation of the EF \cite{G}. Our final conclusions are presented 
in Sec. V.

\section{Two-mode Gaussian states}
 
An undisplaced TMGS is entirely specified 
by its CM denoted by ${\cal V}$ which determines 
the characteristic function of the state
\begin{eqnarray}
\chi_G(x)=\exp{\left(-\frac{1}{2}x^T {\cal V} x \right)},
\label{CF} 
\end{eqnarray}
with $x^T$ denoting a real row vector $(x_1\; x_2\; x_3\; x_4)$. 
${\cal V}$ is a  symmetric and positive $4\times 4$ 
 matrix which has the following block structure:
\begin{eqnarray}
{\cal V}=\left(\begin{array}{cc}{\cal V}_1&{\cal C}\\
 {\cal C}^T&{\cal V}_2 \end{array}\right).
\label{2.26} 
\end{eqnarray}
Here ${\cal V}_1$, ${\cal V}_2$, and ${\cal C}$ are $2\times 2$
matrices. Their entries are second-order correlations of the canonical operators
$q_j=(a_j+a_j^{\dag})/{\sqrt{2}},\; p_j=(a_j-a_j^{\dag})/(\sqrt{2}i)$,
where $a_j $ and $a_j^{\dag}$,  $(j=1,2)$, are the amplitude operators 
of the modes. ${\cal V}_1$ and ${\cal V}_2$ denote the symmetric 
covariance matrices for the individual reduced one-mode STSs \cite{Ma},  
 while the matrix ${\cal C}$ contains the cross-correlations 
between modes. 
The Robertson-Schr\"odinger form of the uncertainty relations for the canonical
variables can be cast as
\begin{equation}{\cal V}+\frac{i}{2}\Omega\geq 0,\;\;
\Omega:=i  (\sigma_2 \oplus \sigma_2),\label{up}
\end{equation}
where we have used the $\sigma_2 $-Pauli matrix. From Eq.\ (\ref{up})
we get the  Sp $(2,\mathbb{R})  \times$ Sp $(2,\mathbb{R})$ invariant inequality
 \cite{Si1,PTH}
\begin{equation}
{\rm det} ({\cal V}+\frac{i}{2}\Omega)={\rm det}{\cal V}-\frac{1}{4}\left({\rm det}{\cal V}_1+
{\rm det}{\cal V}_2+2{\rm det}{\cal C}\right)+\frac{1}{16}\geq 0.
\label{Sp2} 
\end{equation}
A factorized form of the condition \ (\ref{Sp2}) in terms of the symplectic eigenvalues $k_+$ and $k_-$ of the CM,
 \begin{equation}
{\rm det} ({\cal V}+\frac{i}{2}\Omega)
=\left(k_+^2 -
\frac{1}{4}\right)\left(k_-^2-\frac{1}{4}\right)\geq 0,\label{fact1}
\end{equation}
shows that $k_+\geq k_-\geq 1/2.$

According to the separability criterion derived by Simon \cite{Si1}, 
a TMGS is separable if the PT-density operator
${\rho^{PT}}$ describes a Gaussian state. This means that its CM, hereafter denoted by $\tilde{\cal V}$, obeys the uncertainty relation 
\begin{equation}
{\rm det} (\tilde{\cal V}+\frac{i}{2}\Omega)
=\left(\tilde{k}_+^2 -
\frac{1}{4}\right)\left(\tilde{k}_-^2-\frac{1}{4}\right)\geq 0,\label{fact2}
\end{equation}
which is equivalent to the condition  $\tilde{k}_-\geq 1/2.$
\subsection{Scaled standard states}
Following Refs.\cite{Duan,Si1} we define an 
equivalence class of locally similar TMGSs. The states
 belonging to this class have the same {\em amount of 
entanglement} and {\em a scaled standard form} of their 
CM's. These are  four-parameter and two-variable matrices \cite{Duan}:
\begin{eqnarray}{\cal V}(u_1,u_2) & =&\left(\begin{array}{cccc}b_1 u_1 &0& 
c\sqrt{u_1u_2} &0\\
0& b_1/u_1&0&d/\sqrt{u_1u_2}\\ c\sqrt{u_1u_2}  &0&b_2 u_2&0\\0
&d/\sqrt{u_1u_2}&0&b_2/u_2 \end{array}\right)\nonumber\\
&& (b_1\geq 1/2,\;
 b_2\geq 1/2).\label{tri}
\end{eqnarray} 
We denote by 
$\rho^{(I)}$ the Gaussian density operator whose 
CM is 
\begin{eqnarray}{\cal V}_{I}:={\cal V}(1,1).\end{eqnarray}
In fact the CMs \ (\ref{tri}) are obtained by applying  to  ${\cal V}_{I}$
 two independent one-mode squeeze transformations
\begin{equation}{ S}={ S}_1 \oplus { S}_2,\;\;{S}
\in{\rm Sp}(2,\mathbb{R})\times {\rm Sp}(2,\mathbb{R}),\label{sc}\end{equation}
with the squeeze factors $u_1, u_2$. 
${\cal V}_{I}$ was called {\em the standard form I of the
 CM } for this equivalence class  \cite{Duan}.
 There is an obvious  one-to-one correspondence  between the set
 of the four standard-form 
parameters $b_1, \; b_2, \;c, \;d$ appearing as entries in 
${\cal V}_{I}$ and the set of the 
  Sp$(2,\mathbb{R}) \times $ Sp$(2,\mathbb{R})$ 
 invariants ($\det {\cal V}_1, \; \det {\cal V}_2, \; \det {\cal C}$, 
and $\det {\cal V}$). 
According to Simon \cite{Si1}, entangled TMGSs should have a negative $d$ parameter.

Among the scaled standard states, there is another important 
one  introduced and discussed by Duan {\em et al.} \cite{Duan}: the TMGS for 
which the separability and classicality conditions coincide \cite{PTu1}. 
Let us denote its CM by ${\cal V}_{II}={\cal V}(v_1,v_2)$ and term it
 {\em the 
standard form II}. The scaling factors $v_1,v_2$
 satisfy the algebraic 
system   
\begin{eqnarray}\frac{{b_1}(v_1^2-1)}{2 b_1-v_1}=\frac{{b_2}(v_2^2-1)
}{2 b_2-v_2}
\label{f4a}\end{eqnarray}
\begin{eqnarray}b_1b_2(v_1^2-1)(v_2^2-1)=(c v_1v_2-|d|)^2.
\label{f4b}\end{eqnarray} 
Although the solution of the system \ (\ref{f4a})--\ (\ref{f4b})
for an arbitrary TMGS
arises finally from a still unsolved eighth-order one-variable algebraic
 equation, it is possible to find it
 for some particular useful TMGSs.

\subsection{Symmetric TMGSs}
When having  
$\det{\cal V}_1= \det{\cal V}_2$ 
we are dealing with {\em symmetric} TMGSs. The standard parameters 
of the CMs for entangled
 symmetric TMGSs we are considering in the following are
 denoted as
$b_1=b_2=:b,\;c\geq|d|,\; d=-|d|$. We shall review bellow several
useful properties of these states.  

  The symplectic eigenvalues of the CM are
\begin{equation}k_+=\sqrt{(b-|d|)(b+c)},\;\;k_-=\sqrt{(b+|d|)(b-c)}.\label{simp}
\end{equation}
 Similarly, for the PT-density operator we find 
\begin{equation}\tilde{k}_+=\sqrt{(b+|d|)(b+c)},\;\;\tilde{k}_-=\sqrt{(b-|d|)(b-c)}.\label{simpPT}
\end{equation}
leading to the  separability condition \cite{Duan}
\begin{equation}(b-|d|)(b-c)-\frac{1}{4}\geq 0. \label{c2}\end{equation}

Further, Eqs.\ (\ref{f4a}) and \ (\ref{f4b}) can be solved for a symmetric TMGS.
We readily get the squeezed factors in the standard form II:
\begin{equation}v_1=v_2=\sqrt{\frac{b-|d|}{b-c}}. \label{v12}\end{equation}
Note also that the CM of any symmetric TMGS can be diagonalized with a beam-splitter
transformation.
The optical effect of a a lossless beam splitter is described by the {\em 
wave mixing operator} \cite{bonny,leo}
\begin{equation}
B(\theta, \phi)=\exp{\left[-\frac{\theta}{2}(
{\rm e}^{i\phi}a_1^{\dag}a_2-{\rm e}^{-i\phi}a_1a_2^{\dag})\right]}
 \label{bs1}\end{equation}
with $\theta\in[0,\pi],\;\;\;\phi\in(-\pi,\pi].$
Transformation of an arbitrary CM is governed by a
 $4\times4$ symplectic and orthogonal 
matrix
${ M}(\theta,\phi)\in{\rm S O(4)}\cap {\rm Sp}(4,\mathbb{R})$
\begin{equation}{\cal V}^{(B)}= { M}^T {\cal V} { M},
\label{bs2}\end{equation}
where the superscript $T$ stands for transpose.
Explicitly we get
\begin{eqnarray}{M}(\theta,\phi)=\left(\begin{array}{cc}
\cos \frac{\theta}{2}\, I_2& -
\sin \frac{\theta}{2}R(\phi)\\
\sin \frac{\theta}{2}R(-\phi)
&\cos \frac{\theta}{2}\, I_2\end{array}\right),\label{bs3}
\end{eqnarray}
where $I_2$ is the $2\times 2$ identity matrix and 
$R(\phi)$ is the $2\times 2$ rotation matrix
\begin{eqnarray}{R}(\phi)=\left(\begin{array}{cc}\cos \phi&-\sin \phi\\
\sin \phi&\cos\phi\end{array}\right).\label{bs4}
\end{eqnarray}
Now, the CM of a symmetric equally scaled standard state ($u_1=u_2=u$)
  has the nice property of being diagonalized by a beam-splitter transformation
having the angles $\phi=0$ and $\theta=\pi/2$. By applying
 Eq.\ (\ref{bs2}) via Eqs.\ (\ref{bs3}) and \ (\ref{bs4}) we get
\begin{equation}{\cal V}^{(B)}(u,u)={\rm diag}\left[(b+c)u,\frac{b-|d|}{u},(b-c)u,
\frac{b+|d|}{u}\right]. 
\label{di0}\end{equation}
In the particular case of a symmetric TMGS having its CM
 in the standard form II
  the congruent matrix \ (\ref{di0}) reads
\begin{equation}{\cal V}^{(B)}_{II}
={\rm diag}\left[(b+c)\sqrt{
\frac{b-|d|}{b-c}}
,\tilde{k}_{-},\tilde{k}_{-},(b+|d|)\sqrt{\frac{b-c}{b-|d|}}\right].
\label{di2}
\end{equation}

\section{Defining Gaussian entanglement}

We follow now the idea of Vedral and co-workers \cite{Ved}
 and characterize the degree of inseparability of a TMGS by its Bures distance to the set 
of 
 all separable TMGSs of the given system. The original and rigorous proposal
 in Ref.\cite{Ved} is thus modified by restricting
the set of all separable states to a  relevant one
identified by a separability criterion. As for the continuous-variable 
two-mode systems a separability 
criterion was proved for only TMGSs \cite{Si1,Duan}, we find  natural
to use the separable TMGSs as reference set when defining an 
entanglement measure for a symmetric TMGS. 
\subsection{Properties of Uhlmann fidelity}
The virtues of
the Bures distance  \cite{Bures} as a measure of entanglement were 
first revealed in Ref.\cite{Ved}. In our paper \cite{PTH1}, we took advantage
of having derived an explicit  formula for the Uhlmann fidelity \cite{Uhl,Jo}
 between 
two-mode squeezed thermal states and gave the first 
Gaussian amount of entanglement measured by Bures distance. 
Notice that
the Uhlmann fidelity is tightly related to the Bures metric: 
\begin{equation}
d_{B}(\rho,\sigma):=[2-2\sqrt{{\cal F}(\rho,\sigma)}]^{1/2}. 
\label{db}
\end{equation}
In Eq.\ (\ref{db}) ${\rho}$ and ${\sigma}$ are density operators 
acting on a Hilbert space ${\cal H}$ and the function 
${\cal F}(\rho,\sigma)$ is the Uhlmann {\it fidelity} of the two states\cite{Uhl,Jo}: 
\begin{eqnarray}
{\cal F}(\rho, \sigma)=\left\{{\rm Tr}[(\sqrt{\rho}\sigma\sqrt{\rho})
^{1/2}]\right\}^2 \label{3} 
\end{eqnarray}
 
Some of the  remarkable general properties  of the fidelity \cite{Uhl,Jo,F}
 listed bellow will be explicitly used in the rest of the paper.

\begin{description}
\item {\bf P1}  $0\leq {\cal F}(\rho, \sigma)\leq 1,\;\;\;$ and\;\;\; 
${\cal F}(\rho, \sigma)=1\;\;\;$ if and only if \;\;$\rho=\sigma$.
\item {\bf P2}  ${\cal F}(\sigma, \rho)={\cal F}(\rho, \sigma),\;\;\;$ (symmetry).
\item {\bf P3}  ${\cal F}(\rho, \sigma)\geq {\rm Tr} (\rho \sigma);\;\;\;$

if at least one of the states is pure, Eq.\ (\ref{3}) reduces to 
the usual transition probability ${\rm Tr}(\rho \sigma)$, {\it i.e.,} 
if ${\rho}$ or/and ${\sigma}$ is pure, then ${\cal F}(\rho, \sigma)=
{\rm Tr} (\rho \sigma)$.
\item  {\bf P4}   ${\cal F}(U \rho U^{\dag}, U \sigma U^{\dag})=
{\cal F}(\rho, \sigma),\;\;\;$ (invariance under unitary transformations).
\item {\bf P5}  ${\cal F}(\rho_1 \otimes \rho_2, \sigma_1 \otimes \sigma_2)=
{\cal F}(\rho_1, \sigma_1){\cal F}(\rho_2, \sigma_2),\;\;\;$ 
(multiplicativity).
\end{description}

It was also proved that the fidelity cannot decrease
under local general 
measurements and classical communications. This property 
of fidelity is important in  properly defining 
a measure of entanglement.
\subsection{Defining Gaussian entanglement by Bures metric}
Let us denote in the following by $\rho_s$ the density operator of an inseparable
 symmetric 
TMGS whose CM, Eq.\ (\ref{2.26}), has the following structure:
\begin{eqnarray}
{\cal V}_1={\cal V}_2=\left(\begin{array}{cc} b u&0\\
0&b/u\end{array}\right),\;
{\cal C}=\left(\begin{array}{cc} c u&0\\
0&-|d|/u\end{array}\right),\nonumber \\
\left(b\geq\frac{1}{2} \right).
\label{red} 
\end{eqnarray} 
Equation \ (\ref{red}) describes the CM of a symmetric scaled standard state
with equal
local squeeze factors. 
For later convenience, we define 
{\em  the amount of Gaussian entanglement} of the entangled state $\rho_s$
\begin{equation}
E_0(\rho_s):=\min_{{\rho}^{\prime} \in {\cal D}_0^{sep}}\frac{1}{2} 
d^2_B(\rho_s, {\rho}^{\prime})=1-\max_{{\rho}^{\prime} \in {\cal D}_0^{sep}}
\sqrt{{\cal F}(\rho_s, {\rho}^{\prime}}). 
\label{set} 
\end{equation}
In Eq.\ (\ref{set}) we have introduced 
the set ${\cal D}_0^{sep}$
of {\em all separable scaled standard TMGSs} which is included in
the set of all separable TMGSs displaying the property (separability threshold) \cite{PTH1}
\begin{equation}
  \tilde{k^{\prime}}_-=1/2.
\label{s}\end{equation}
Our task is to maximize the fidelity between the entangled symmetric
TMGS  $\rho_s$ 
and a state ${\rho}^{\prime}\in{\cal D}_0^{sep}$.
Obviously, as the inseparability does not depend on 
local operations we have 
\begin{equation}
E_0(\rho_s)=E_0(\rho_I).\label{set1} 
\end{equation} 

\section{Evaluating Gaussian entanglement}
\subsection{Results on the fidelity between two Gaussian states}
 Evaluation of
the fidelity between one-mode Gaussian states  
was possible by taking advantage of the exponential form of their
density operators. According 
to Ref.\cite{TS}, the fidelity between the undisplaced one-mode Gaussian states $\sigma_1$ and $\sigma_2$ is 
\begin{eqnarray}{\cal F}(\sigma_1,\sigma_2)=[(\Delta+\Lambda)^{1/2}-\Lambda^{1/2}]^{-1}\label{onem}\end{eqnarray}
with
\begin{eqnarray}\Delta= \det({\cal V}_{\sigma_1}+{\cal V}_{\sigma_2}),\nonumber\\
 \Lambda=4[\det({\cal V}_{\sigma_1})-\frac{1}{4}] [\det({\cal V}_{\sigma_2})-\frac{1}{4}].
\label{fom}
\end{eqnarray}
A main consequence of having an analytic formula for the fidelity 
was to define and calculate a
degree of nonclassicality  for one-mode Gaussian states \cite{PTHlet}. 
 The explicit formula of the fidelity was then used to quantify the
 accuracy of  teleportation of mixed  one-mode  Gaussian states 
through a Gaussian channel in  Refs.\cite{PTH03,ban}.

 It appears that
 our paper  \cite{PTH1} co-authored with Scutaru was the only one to give and exploit 
an explicit formula for the fidelity of two TMGSs. After
  evaluating  the fidelity
 between two two-mode STSs, we obtained an explicit expression
for a properly defined Gaussian  amount of entanglement of a STS.
Recently, we 
have shown that the fidelity between two TMGSs having the density
operators $\rho_1$ and $\rho_2$ is determined by the properties of the 
non-Hermitian Gaussian operator $\rho_1\rho_2$ \cite{PTu}. Here we  want 
to deal with the Bures entanglement \ (\ref{set})
of a symmetric TMGS.  Fortunately,  the details of
 the explicit general formula of the fidelity between two TMGSs
are not
necessary here. Instead,  we have to use
the following property which we have  proved by using the general formula for 
the fidelity between scaled standard states \cite{PTu}:
The closest separable scaled standard state to a given symmetric scaled 
standard state having equal local squeeze factors  $u_1=u_2=u$  is a 
a similar symmetric scaled standard state observing the threshold 
condition \ (\ref{s}).
 Therefore, the amount of Gaussian entanglement for a symmetric
 TMGS can be calculated in a simpler way, 
because the separable reference set  ${\cal D}_0^{sep}$ used in Eq.\ (\ref{set})
is in fact 
restricted to its subset of 
equally squeezed symmetric states. 
 
By using property {\bf P4} of the fidelity for the beam--splitter operator 
 $B(\pi/2, 0)$, Eq.\ (\ref{bs1}), we first write the fidelity between $\rho_s$ 
and any symmetric scaled 
standard state ${\rho}^{\prime}$ having equal local squeeze factors: 
\begin{eqnarray}{\cal F}(\rho_s, {\rho}^{\prime})=
{\cal F}[B(\frac{\pi}{2}, 0){\rho_s}B^{\dagger}(\frac{\pi}{2}, 0), B(\frac{\pi}{2}, 0){\rho}^{\prime}B^{\dagger}(\frac{\pi}{2}, 0)].\nonumber\\
\label{tri2}.
\end{eqnarray}
According to Eq.\ (\ref{di0}), 
the transformed density operators describe  two-mode product states:
\begin{eqnarray}B(\pi/2, 0){\rho_s}B^{\dagger}(\pi/2, 0)=\sigma_1\otimes 
\sigma_2\end{eqnarray}
and respectively
\begin{eqnarray}B(\pi/2, 0){\rho}^{\prime}B^{\dagger}(\pi/2, 0))
=\sigma_1^{\prime}\otimes 
\sigma_2^{\prime}.\label{trsep} \end{eqnarray}
Here the one-mode states $\sigma_1$ and respectively $\sigma_2$ have the CMs
\begin{eqnarray}
{\cal V}_{\sigma_1}(u,u)={\rm diag}[(b+c)u,(b-|d|)/u],\nonumber\\
{\cal V}_{\sigma_2}(u,u)={\rm diag}[(b-c)u,
(b+|d|)/u]. \label{di0B}\end{eqnarray}
Similarly, the transformed separable density operator \ (\ref{trsep})
has the CMs of its one-mode reductions 
\begin{eqnarray}
{\cal V}_{\sigma_1^{\prime}}(u^{\prime},u^{\prime})={\rm diag}
[(b^{\prime}+c^{\prime})u^{\prime},(b^{\prime}-|d^{\prime}|)/u^{\prime}],\nonumber\\
{\cal V}_{\sigma_2^{\prime}}(u^{\prime},u^{\prime})=
{\rm diag}[(b^{\prime}-c^{\prime})u^{\prime},
(b^{\prime}+|d^{\prime}|)/u^{\prime}]. \label{di0Bprime}\end{eqnarray}
However, the parameters appearing in the above equation are related by the 
threshold separability condition \ (\ref{s}). 
Now we can apply the
multiplicativity property {\bf P5} of the fidelity and reduce the evaluation
of fidelity  to a single-mode problem \cite{Oliveira}:
\begin{eqnarray}{\cal F}(\rho_s, {\rho}^{\prime})=
{\cal F}(\sigma_1,{\sigma}_1^{\prime})
{\cal F}(\sigma_2,{\sigma}_2^{\prime}).\label{mult}\end{eqnarray}

\subsection{Explicit evaluation}
Maximization of the product-fidelity \ (\ref{mult}) with respect 
to the parameters 
of the CM \ (\ref{di0Bprime}) is still a complicated problem. At this point
 we choose to use the 
 standard form II of the covariance matrix, Eq.\ (\ref{di2}), for
 the given entangled state $\rho_s.$ The separability threshold condition 
\ (\ref{s}) 
is manifestly satisfied when  
the closest separable state ${\rho}^{\prime}$ is in the standard form II, too.
We are now left to maximize the product of the one-mode fidelities 
between states described by the CMs
\begin{eqnarray}
{\cal V}_{\sigma_1}={\rm diag}\left[(b+c)\sqrt{
\frac{b-|d|}{b-c}}
,\tilde{k}_{-}\right],\nonumber\\
{\cal V}_{\sigma_1^{\prime}}={\rm diag}\left[2 (b^{\prime}+c^{\prime})(b^{\prime}-|d^{\prime}|), 
\frac{1}{2}\right],
 \label{di0B1}\end{eqnarray}
and 
\begin{eqnarray}
{\cal V}_{\sigma_2}={\rm diag}\left[\tilde{k}_{-},(b+|d|)\sqrt{\frac{b-c}{b-|d|}}\right],\nonumber \\
{\cal V}_{\sigma_2^{\prime}}=
{\rm diag}\left[\frac{1}{2},2(b^{\prime}+|d^{\prime}|)(b^{\prime}-c^{\prime})\right].
 \label{di0B2}\end{eqnarray}
Application of Eqs.\ (\ref{onem}) and \ (\ref{fom}) for one-mode fidelities shows that the product \ (\ref{mult}) is a function of only two independent variables \cite{com1}:
 \begin{eqnarray} x:=(b^{\prime}+c^{\prime})(b^{\prime}-|d^{\prime}|),\;\;
y:=(b^{\prime}+|d^{\prime}|)(b^{\prime}-c^{\prime}).\label{newvar}\end{eqnarray}
In this way 
 a considerable 
simplification of  maximization procedure is obtained.
 We easily find that the maximal fidelity
\begin{eqnarray}\max_{{\rho}^{\prime} \in {\cal D}_0^{sep}}
{\cal F}(\rho_s, {\rho}^{\prime})
=\frac{2 \tilde{k}_-}{\left(\tilde{k}_-+1/2\right)^2},
\label{maxf}\end{eqnarray}
is reached when the following conditions are met:
\begin{eqnarray}
x_{max}-\frac{1}{4}=\frac{1}{2\tilde{k}_-}\left(k_+^2-\frac{1}{4}
\right)
\nonumber\\
y_{max}-\frac{1}{4}=\frac{1}{2\tilde{k}_-}\left(k_-^2-\frac{1}{4}
\right).
\label{varm}\end{eqnarray}
Remark that the maximal fidelity is a function of
 only $\tilde{k}_-$ while the parameters
$x_{max}$ and $y_{max}$ depend on the symplectic eigenvalues \ (\ref{simp}) as well.
We can write now the final expression for the Gaussian
 entanglement measured by the Bures metric,
Eq.\ (\ref{set}),
\begin{eqnarray}
&&E_0(\rho_s)=\frac{(1-\sqrt{2\tilde{k}_-})^2}{2\tilde{k}_-+1}
,\;\;\tilde{k}_-<1/2, \nonumber\\&&
E_0(\rho_s)=0,\;\;\tilde{k}_- \geq 1/2.\label{fin}
\end{eqnarray}

Recall that in  Ref.\cite{G} it was proved that the {\em exact}
 EF of a symmetric TMGS
is a function of only $\tilde{k}_-$. Our result \ (\ref{fin}) obtained 
in the Gaussian approach of Bures-metric entanglement is thus in agreement with
the exact EF. Accordingly, one could assume that, in terms of the Bures metric, the  closest separable state to an entangled
symmetric TMGS is also a symmetric TMGS. 
\subsection{The closest separable state}
It is instructive to determine the parameters 
$b^{\prime\prime}, c^{\prime\prime},d^{\prime\prime}=-|d^{\prime\prime}|$ 
of the closest separable state. To this end we use
 Eqs.\ (\ref{s}) and \ (\ref{varm}) and get
\begin{eqnarray}
&&(b^{\prime\prime})^2=\frac{1}{4{\tilde{k}_-}^2}
\left[{\tilde{k}_-}+k_+^2-\frac{1}{4}\right]\left[{\tilde{k}_-}+
k_-^2-\frac{1}{4}\right]
\nonumber \\&&
c^{\prime\prime}=\frac{1}{2\tilde{k}_-} \sqrt{
\frac{\tilde{k}_-+k_-^2-1/4}{\tilde{k}_-+k_+^2-1/4}}\left(k_+^2-1/4\right)\nonumber \\&&
|d^{\prime\prime}|=\frac{1}{2\tilde{k}_-} \sqrt{\frac{\tilde{k}_-+k_+^2-1/4
}{\tilde{k}_-+k_-^2-1/4}}\left(k_-^2-1/4\right).\label{prm}\end{eqnarray}
Equations \ (\ref{prm}) tell us that the parameters of the 
 closest separable state
 to a symmetric TMGS  are determined by the symplectic eigenvalues
 $k_-$ and $k_+$
of its CM and by the smallest symplectic eigenvalue $\tilde{k}_-$ of the CM ${\cal V}$
of the PT-density operator.

\section{Conclusions}
The intricate 
 expressions \ (\ref{prm}) give one an idea about the considerable 
simplification we introduced in
the maximization procedure of the fidelity in two ways: first by applying  
its property {\bf P4} under the beam-splitter transformation, second by
 considering the given state with its CM in the standard form II. Notice that
 this form of the CM is involved in giving an inseparability criterion
for a TMGS \cite{Duan}. 
Our result for the Gaussian degree of entanglement measured by Bures distance
depends only on the smallest symplectic eigenvalue
 of the covariance matrix of the PT-density operator. Thus, it is 
in agreement with the exact EF found in Ref.\cite{G}. One could therefore conjecture that the closest separable state to an entangled symmetric TMGS in terms of Bures metric is a Gaussian.  This is not the case with other distance-type measures of entanglement such as relative entropy \cite{Ved} for which it is known  that, even for pure states, the closest separable state to a Gaussian one is non-Gaussian.

\section*{Acknowledgments}
We are grateful to M. S. Kim for bringing Ref.[7] to our attention. This work was supported by the Romanian 
Ministry of Education and Research
through Grants No. CEEX 05-D11-68/2005 and No. IDEI-995/2007 for the University of Bucharest.


\begin{thebibliography}{99}
\bibitem{com}These two Gaussian measures of entanglement for a STS are interestingly
depending on the difference between the two-mode squeeze parameter $r$ and its 
value $r_s$ defining  the separability threshold.
\bibitem{PTu1}  Generally,
the  classicality condition (existence of a well--behaved $P$ representation)
 is stronger than the separability one. See our
paper \cite{PTH} for a more detailed analysis
 on this issue.
\bibitem{com1} The new variables $x$ and $y$ of maximization are in fact the 
squared symplectic eigenvalues of the symmetric separable state $\rho^{\prime}$.
\end{thebibliography}
\end{document}